\documentclass[12pt, preprint]{aastex}

\begin{document}

\title{HST/NICMOS Observations of NGC 1333: The Ratio of Stars to Sub-stellar Objects}

\author{Julia Greissl}
\affil{Steward Observatory, University of Arizona, Tucson, AZ 85721}
\email{jgreissl@as.arizona.edu}

\author{Michael R. Meyer}
\affil{Steward Observatory, University of Arizona, Tucson, AZ 85721}
\email{mmeyer@as.arizona.edu}

\author{Bruce A. Wilking}
\affil{Department of Physics and Astronomy, University of St.Louis \\
  1 University Blvd., St. Louis, MO 63121}

\author{Tina Fanetti}
\affil{Department of Physics and Astronomy, University of St.Louis \\
  1 University Blvd., St. Louis, MO 63121}

\author{Glenn Schneider}
\affil{Steward Observatory, University of Arizona, Tucson, AZ 85721}

\author{Thomas P. Greene}
\affil{NASA AMES Research Center, Moffet Field, CA 94035}

\author{Erick Young}
\affil{Steward Observatory, University of Arizona, Tucson, AZ 85721}

\begin{abstract}
We present an analysis of NICMOS photometry and low-resolution grism
spectroscopy of low-mass stars and sub-stellar objects in the young 
star-forming region NGC 1333. 
Our goal is to constrain the ratio of low-mass stars to sub-stellar objects 
down to 20 M$_{Jup}$ in the cluster as well as constrain the cluster IMF down 
to 30 M$_{Jup}$ in combination with a previous survey of NGC 1333 by Wilking et al. 
Our survey covers 4 fields of 51.2'' x 51.2'', centered on 
brown dwarf candidates previously identified in Wilking et al. 
We extend previous work based on the use of a water vapor index 
for spectral typing to wavelengths accessible with NICMOS on the HST. 
Spectral types were derived for the 14 brightest objects in our fields, 
ranging from $\leq$M0 - M8, which at the age of the cluster (0.3 Myr) corresponds to
 a range in mass of $\geq$0.25 - 0.02 M$_{\odot}$.
In addition to the spectra, we present an analysis of the color-magnitude 
diagram using pre-main sequence evolutionary models of D'Antona \& Mazzitelli.  
Using an extinction-limited sample, we derive
the ratio of low-mass stars to brown dwarfs. Comparisons of the observed ratio
to that expected from the field IMF of Chabrier indicate that 
the two results are consistent. We combine our data with that of 
Wilking et al. to compute the ratio of intermediate-mass stars 
(0.1 - 1.0 M$_\odot$) to low-mass objects (0.03 - 0.1 M$_\odot$) in the cluster.
We also report the discovery of a faint companion to the previously confirmed brown 
dwarf ASR 28, as well as a possible outflow surrounding ASR 16. If the faint 
companion is confirmed as a cluster member, it would have a mass of $\approx$ 
5 M$_{Jup}$ (mass ratio 0.15) at a projected distance of 350 AU, similar 
to 2MASS 1207-3923 B.
\end{abstract}


\section{Introduction}

The shape of the Initial Mass Function (IMF) and its connection to the initial 
physical conditions in molecular clouds remains one of the fundamental 
questions in star formation. More specifically: Does the young cluster IMF mimic 
the integrated field star IMF even to very low masses? Is there a low-mass cut-off 
to the sub-stellar mass function? 

The IMF over the full range of stellar masses has been extensively studied 
in the past decades starting with \citet{ms79}, as well as updates by 
\citet{ktg93} and \citet{ch03}. It 
is generally accepted that young clusters exhibit a Salpeter-like IMF above
1 M$_{\odot}$, flattening out towards low-mass stars. Studies of the IMF
 have been concentrated on young clusters, because they
offer several beneficial characteristics. Their populations are less likely to 
have 
undergone significant dynamical mass segregation compared to older clusters,
 which means that a small field can 
yield a sample representative of the whole cluster. In addition, young low mass 
stars are still
located significantly above the main sequence and are thus several times brighter than
their main sequence counterparts. Lastly, young 
clusters are fairly compact, meaning that 
they occupy a small area in the sky as well as often being situated in front
of a large amount of extinction. This minimizes both foreground
contamination by field stars and background contamination by giants. \\

Recent studies of galactic young clusters
have been extended down to the hydrogen-burning limit to a 
distance of 1 kpc (\citet{luh98}; \citet{hi97}; \citet{car97}). 
These studies suggest that the IMF is universal 
above 0.1 M$_\odot$ \citep{mey00}. 
The shape of the IMF in the sub-stellar regime is less well constrained and
we are only now beginning to probe it in great detail (\citet{hi00}; \citet{bri02}). 
Several clusters have been studied down to 30 M$_{Jup}$ in both 
photometric and spectroscopic studies, among them the Orion Nebula Cluster (\citet{sl05}), 
IC348 (\citet{na00}; \citet{luh03a}), the Taurus star forming region (\citet{luh04b}; \citet{gu05}), 
the Chameleon I star forming region (\citet{luh04}) and the Pleiades (\citet{mo03}).

Searching for IMF variations at the lowest masses 
can tell us about possible low-mass cut-offs in the IMF, which
might be associated with the minimum mass for opacity limited-fragmentation 
($\approx$ 10 M$_{Jup}$) \citep{re76}.  
Evidence presented by \citet{bri02} (see also: \citet{luh04}) 
showed the Taurus star-forming region to exhibit a dearth of
brown dwarfs, by as much as a factor of 2, when compared to the denser region 
of the Orion Nebula Cluster.
 However, newer studies, covering a larger area of Taurus suggest that this deficiency 
is not as marked as previously reported \citep{luh06}. Nonetheless the low-mass IMF 
remains a strong candidate for variations with cluster characteristics.
This suggests a possible connection between stellar density and the 
shape of the low-mass IMF. NGC 1333 is a young star-forming region intermediate
in density between Taurus and Orion, making it an ideal test bed to search
for variations in the low-mass IMF and explore differences based on star-forming
environment.

The NGC 1333 reflection nebula is part of the Perseus molecular cloud complex
at an estimated distance of 300 pc (\citet{dZ99}; \citet{be02}). The 
proximity of the cluster allows us to study the IMF down to very low masses. 
It was first identified
as a star-forming region in \citet{hr72} (see also: \citet{he74}; \citet{str76}). 
Infrared surveys
have since revealed a large population of low-mass objects 
(\citet{as94}; \citet{la96}; \citet{wil03}). These studies have characterized the 
young stellar population over large areas of the cloud, but have not fully 
probed the very lowest masses because of a lack of sensitivity.   

We have performed a new, deep (J $<$ 21$^m$) near-infrared survey over a 
limited area in NGC 1333 using NICMOS/HST. This study is complete to fainter 
magnitudes than previous studies, thus allowing the characterization of a significant extinction-limited sample 
of brown dwarfs. This enables us to study the population of very low-mass objects in 
detail. We obtained photometry as well as infrared 
spectroscopy. The spectra enable estimates of stellar masses and ages 
for individual objects by comparing their positions in
the H-R Diagram with PMS evolutionary models. The spectra further 
help us adopt models to interpret our flux-limited survey by constraining a 
mass-luminosity relation appropriate for the cluster.  
We construct an extinction limited sample to explore the shape
 of the mass function in the cluster. 

The paper is structured as follows. 
Section 2 describes the observations and data reduction, followed by a 
description of the photometric and spectroscopic analysis in Section 3. 
Section 4 details the results, and places them in context with recent
literature. A summary and conclusions are presented in Section 5.

\section{Observations and Data Reduction}

We obtained images of the young cluster NGC 1333 using camera 3 of NICMOS 
(NIC3) as part of the HST program 9846. Six fields, each 51.2'' x 51.2'' with a 
plate scale of 0.2''pixel$^{-1}$, were obtained between 
2004 Jan 15 and 2005 Aug 11 using 12 HST orbits. In addition to the 
6 NGC 1333 fields we also obtained observations for 7 additional objects. Six of these 
are previously confirmed 
brown dwarfs from older star-forming regions and the field. We obtained these observations
 in order to explore the surface gravity dependence of our 
spectral typing technique. 
The coordinates for each of the fields in addition to the total exposure times 
are listed in Table 1. For the position of our fields with respect to molecular line maps 
of the region please see Figure 1 in \citet{wil03}. Each field was 
observed in F110W, F160W and the grism G141, which cover 0.8 - 1.4 $\micron$ 
(broadband), 1.4 - 1.8 $\micron$ (broadband) and 1.1 - 1.9 $\micron$  
(dispersed), respectively. F110W is roughly
equivalent to J-band and F160W is roughly equivalent to H-band. 
The fields were observed in a 2 x 2 dither pattern, with an offset of 1.75 
pixels. The observations were subdivided into visits, with each visit covering
one field in G141, F160W and F110W in that order. 
Since the grism G141 is slitless, stars that are close in either the
x or the y direction can have overlapping spectra for each individual visit.
To minimize this overlap we observed each photometric and spectroscopic 
field in 3 visits, at roll angles
offset 30 - 75$^o$ and 105 - 150$^o$ between subsequent visits.
We were unable to obtain 3 separate roll angles for the fields N2 and N7 
causing their final photometry to be less deep than that of the other 4 fields. 
For this reason they are only used in the spectroscopy section 
and excluded from the photometric analysis. All 4 of these fields lie in the southern 
half of the cluster and the complete area covered 
by our photometric survey is 2.9 arcmin$^2$. Compared to the combined areas 
of the surveys by \citet{wil03} and \citet{as94}, which surveyed both the 
north and south components of the cluster and covered 79 arcmin$^2$ and 81 arcmin$^2$ 
respectively, our photometric survey thus covers $\approx$ 2\% of the area of 
NGC 1333.
Total integration times for the fields with 3 roll angles, including dithers 
and multiple roll angles were 383 s for F160W, 766 s for F110W and 4608 s for G141 per field.

\subsection{Photometry}

Data reduction was carried out using a combination of IRAF and custom IDL 
routines. The methods described below closely follow those used in 
\citet{liu03}. The images were dark and sky subtracted using combined 
dark 
plus sky frames created with the routine NICSKYDARK in the NICRED package for 
IRAF \citep{mc97}. Cosmic rays and bad pixels were located and removed 
using the routine FULLFITBAM, by searching for discontinuities in the flux
in each pixel over time. After dark and sky subtraction and cosmic ray reduction,
there did not appear to be any bias offsets or ``pedestal effects'' 
between the different quadrants of the NICMOS chip, which is a common problem
experienced with NICMOS (e.g. \citet{liu03}). This is 
because our fields were not very crowded and did not show significant 
nebulosity, as the pedestal is also dependent on the total charge of the 
quadrant. The only field that shows any nebulosity is S2.
Finally, the images were flatfielded using the routine NICFLATTEN and the 
appropriate epoch on-orbit flatfields from the Space Telescope Science Center Institute 
(STScI) website. 
The dithers were then combined using the IDL software IDP3 \citep{ly99}, aligning the 
different frames by their World Coordinate System as well as resampling the 
images by a factor of 2 using a bi-cubic sinc to increase the resolution of the images. 
The three roll angles were combined using IDP3, first rotating the images to align in angle
and then shifting the images to correct for offsets in x and y.
For both of these routines, flux was conserved. 

All sources presented in the Color-Magnitude Diagram (CMD) (Sec. 3.3) were detected in both the F160W and F110W 
filters using the IRAF routine DAOFIND with a 10 $\sigma$ detection threshold. Even 
with the low amount of nebulosity every image still had some spurious detections which 
were removed by visual inspection. We took advantage of the three roll angles for each 
field to help identify false detections. If a source appeared in the same position 
for each roll angle, it was assumed to be genuine. False detections were located at their
position on the chip for only one roll angle. 

Since the fields were not crowded, the photometry was
performed using the IRAF routine APPHOT. The optimal aperture size was 
calculated to be 8 pixels in radius with the sky background measured 
using an annulus between 10 and 12 pixels around each object. Aperture corrections 
were calculated out to 25 pixels using several bright, non-saturated stars and 
applied to the photometry.
The aperture corrections derived were 0.054 $\pm$ 0.008 mag for m110 and 
0.073 $\pm$ 0.008 mag for m160, where m110 and m160 refer to the magnitudes associated
with F110W and F160W, respectively. Errors in the photometry were computed 
using PHOT and agreed with errors computed by comparing photometry derived from
each separate roll angle.  Two objects, ASR 9 and ASR 28, showed close companions. 
For these an aperture radius of 3 pixels was used with appropriate aperture
corrections of 0.32 $\pm$ 0.03 mag for m110 and 0.38 $\pm$ 0.02 mag for m160 and a sky annulus of 
20 to 24 pixels so as to exclude flux from the companion. In addition, one object, ASR 16, 
appears to have extended nebulosity associated with it. Because of this, photometry was 
extracted for this object in the same manner as the binaries to minimize contamination 
from the nebulosity. This source is further discussed in section 4.2. 
A total of 25 unique sources were detected and are listed in Table 2.

NICMOS magnitudes were calibrated using zero-points of 1775 and 1093 Jy and 
were calibrated to the Vega system using the conversions of 2.873 x 10$^{-6}$ and
2.776 x 10$^{-6}$ Jy ADU$^{-1}$ s$^{-1}$ for F110W and F160W respectively. 
Detection limits were assessed using artificial star tests. Artificial 
stars were added to the images in 0.5 magnitude steps using a PSF derived 
from bright, non-saturated sources. The recovery fraction of these stars 
was then computed using the same
detection technique as described above. The 90\% completeness limits were found
to be m160 $\approx$ 20.5$^m$ and m110 $\approx$ 21.0$^m$. The NICMOS 
magnitudes m110 and m160 were transformed to the CIT system, J and H, 
by comparing magnitudes for sources in the survey which also had 2MASS 
photometry (a total of 12 objects). The 2MASS photometry was converted to the 
CIT system using the transformations of \citet{ca00}. A linear regression was 
then performed to obtain color transformations between the HST magnitudes and 
the CIT system. The following color corrections were determined:
\\
\\
$
H = m160 + (0.189 \pm 0.030) + (0.120 \pm 0.025)(m110 - m160) 
$
\\
$
J-H = (0.132 \pm 0.073) + (0.760 \pm 0.060)(m110 - m160) 
$
\\
\\
These color transformations were used to transform the NICMOS magnitudes to CIT
with the J magnitudes calculated as J = (J-H) + H. 

Relative astrometry for each object was determined using the World Coordinate System 
in the header of the NICMOS images together with the XY2SKY routine in the WCSTools 
package \citep{mi02}. The derived astrometry was compared with previously determined 
coordinates from 2MASS and showed no systematic offset. All coordinates presented in 
Table 3 are estimated to be accurate to $\leq$ 1.5''. 

\subsection{Spectroscopy}

The spectroscopic observations were performed with the grism G141, which is 
centered at 1.4 $\micron$, with a wavelength coverage between 
1.2 $\micron$ - 1.9 $\micron$ and a resolution
of $\approx$ 200 per pixel. This covers the water-band feature at 1.4 $\micron$ which 
is expected in the atmospheres of cool objects and is difficult to observe 
from the ground, making HST ideal for these observations. To reduce these data, 
the images were first run through the HST pipelines CalnicA and 
CalnicB. These pipelines perform the dark subtraction, using artificial 
dark frames from the STScI website, as well as cosmic ray reduction. 
The pipelines were not used to
co-add either the dithers or the roll angles, because this did not improve the
quality of the spectra over extracting them from individual images. This means 
that for each object in the NGC 1333 sample there can be a maximum of 12 extracted spectra.
The spectra were extracted 
using a custom IDL routine NICMOSlook \citep{pi98}, which was designed to 
deal specifically with NICMOS grism data. NICMOSlook reduces spectra
in a standard way, by tracing the spectrum across the chip and summing
 the flux in the spatial dimension as well as subtracting a background
region set by the user at each wavelength. 
To extract spectra, NICMOSlook needs a spectroscopic as well as a photometric
image, to perform the wavelength calibration. The photometric image gives the 
true location of the object's position, 
which then gives the zero point for the wavelength of the spectrum. The photometric image used was 
the F160W image because its central wavelength 
is closer to the central wavelength of G141. The spectra were extracted with 
an extraction width of roughly 2 x FWHM
of the spectra, in general $\approx$ 4 pixels. The background
size was varied depending on the crowding of the field. NICMOSlook also 
flatfields the spectra during the extraction. This is necessary because the
quantum efficiency of the detector changes with both wavelength and position.
For this reason the spectra cannot be flatfielded before extraction. 
NICMOSlook flatfields the spectra using a set of narrowband
flatfield images, which are chosen according to the date of the
observations, from which the software constructs a 3-dimensional calibration datacube.

Signal-to-noise ratios (SNR) were estimated for all extracted spectra in two ways. 
First, a 3rd order polynomial was fitted to the combined average spectrum in the 
spectral region between 1.5 - 1.8 $\micron$ and the RMS of the difference between the 
observed continuum 
and the fit per pixel calculated. Second, a SNR was calculated for each individual
pixel in the spectrum, by estimating the signal as the mean of each pixel and
the noise as the error in the mean from the ensemble of individual spectra for each 
source. Since there are only four independent extractions
 for some of the spectra, the error in the mean is not necessarily a high fidelity
estimate for the noise in each pixel. This means that the SNR estimated in
this way may overestimate the real SNR of the spectra. The SNR of the spectrum 
is then the average of the SNR for all the pixels.

\subsection{Spectral Standards}

To help calibrate our method for spectral typing we obtained spectra of 7 objects 
with previously known spectral types (see Table 1), six of which are brown dwarfs. 
These objects are older than the NGC 1333 sources and were observed to 
explore the surface gravity dependence in our spectral typing technique. 
 Of the 7 objects, only 4 had high enough SNR to be included in the spectral sequence. They span a spectral 
range from M8 to L7. In addition we extracted spectra from archival HST grism data 
(Program IDs \#7322 and \#7830; see also \citet{na00}) of old low-mass field dwarfs 
with known spectral types. They span a spectral range from M0 - L5. The spectra are 
plotted in Figures 1 and 2. All standards used in the spectroscopic analysis are listed 
in Table 3.

\subsection{NGC 1333 Spectra}

Spectra were extracted for a total of 14 objects in NGC 1333 from 6 fields. 
Depending on the position of the spectra on the chip, such as proximity to 
other objects and position with respect to the edge of the chip, between 4 
and 12 spectra were extracted for each source. 
The spectra were then combined in a sigma-clipped way, excluding pixels which deviated 
by more than three sigma. 
For the NGC 1333 objects 
the SNRs ranged from $\approx$ 30 to $\approx$ 100 and the different SNR estimates 
agreed well with each other. We compared these values with the expected SNR 
from the STScI exposure time calculator. In general these theoretical SNR tended 
to be higher than those estimated from the real spectra. We chose to adopt the SNR 
estimates from the polynomial fits.
Spectra of all NGC 1333 objects are plotted in Figure 3.  

\section{Analysis}

\subsection{Spectroscopic Analysis}

Spectral typing was performed using a water vapor absorption index centered
on the water band at 1.4 $\micron$. This band is
sensitive to temperature in cool stars later than M0 \citep{jo94} but not 
strongly dependent
on surface gravity (\citet{go03}; \citet{wil03}). The method described here 
is similar to the one detailed in \citet{wgm99} and \citet{wil03}. We define 
a Q index that is independent 
of reddening, using the reddening law of \citet{co81}: 

$Q=(F1/F2)(F3/F2)^{0.567}$\\
\\
where F1, F2 and F3 are the averaged fluxes in the bands 1.30 - 1.35 $\micron$, 
1.40 - 1.45 $\micron$ and 1.65 - 1.70 $\micron$ respectively. The index is 
independent of reddening because it includes only a ratio of flux bands which 
have been scaled by an exponent directly related to the applicable reddening law. 
This accounts for any inherent slope in the spectrum due to interstellar reddening 
and corrects for it. 
To optimize our spectral typing technique we chose among several different 
combinations of flux bands and determined which fit of Q versus spectral type for 
our spectral standards showed the lowest combined error in 
slope and intercept and at the same time showed the strongest evolution 
with spectral type. 
This index was chosen because of its tight fit and its strong evolution with spectral 
type, as it directly measures the evolution in the depth of the water 
band feature at 1.4 $\micron$. The evolution of Q with spectral type for the NGC 1333 
objects is shown in Figure 3. The uncertainty in Q was estimated by calculating Q for each individual spectrum of an object 
and then deriving the error in the mean of those values from the multiple spectra
which were available for each object.
To calibrate the relation between Q and spectral type, a weighted linear
fit of Q vs. spectral types for the old field dwarf standards and the young 
confirmed brown dwarfs was performed. Together these
standards have spectral types ranging between M0 - L7 (see Table 3). Spectra with types 
earlier than M0 were excluded because they show no water band absorption. 
This fit gives the relation:

$MV(subclass) = (-5.04 \pm 0.31)Q + (-2.22 \pm 0.47)$ \\
\\
The fit, together with the Q values for the field dwarfs, is plotted in Fig. 4. 

We tried to assess the surface gravity dependence of 
the Q index by deriving a fit of Q vs. spectral type for the young brown 
dwarfs with known spectral types observed as part of our program. Field stars 
in general have log(g) = 5.0 - 5.5, while younger PMS stars have lower surface 
gravities between log(g) = 3.0 - 4.2 \citep{go03}. Thus it is important
to explore whether our spectral typing technique is valid across a range of
log(g). We performed a fit using only the young confirmed brown dwarfs and compared it 
to the fit derived using the field dwarf standards. Comparing these two fits, we estimate 
that assigning spectral types using the old M dwarfs
might underestimate their temperature by $\approx$ 200 K (too late by one subclass). 
This means the temperatures we derive are lower limits, similar to previous findings 
by \citet{go03} and \citet{wil03}. We only have high SNR spectra 
for 3 young brown dwarfs however, and so these are preliminary results at best.

Table 4 
shows the derived spectral types for the NGC 1333 objects.
The spectral types derived for the NGC 1333 objects range from $<$M0 to M8. 
Stars with no water band absorption features were assigned a spectral type of $<$M0.
Uncertainties in Q were between $\pm$0.01 - 0.17, which leads to 
uncertainties in spectral types of approximately one subclass, comparable to the 
potential error in our calibration mentioned above. From this relation we classify seven 
objects with spectral types of M6-M8 which are likely brown dwarfs. Three of the objects for 
which we have derived spectra, ASR 24, ASR 15 and ASR 17, have previous spectral classifications
using K-band H$_2$O absorption bands \citep{wil03}. Both derived spectral types are listed in
 Table 4 and agree to within the errors. While we did not detect cooler young brown dwarfs 
than \citet{wil03}, we were able to observe similar objects with greater obscuration.       
Alternative methods of the spectroscopic reduction as well as a method of 
spectral typing are also addressed in detail in \citet{fa05}. 

\subsection{H-R Diagram}

To estimate the age of the cluster we compared the positions of the NGC 1333 PMS
objects in the H-R Diagram to theoretical evolutionary
tracks by \citet{dm97} (hereafter DM98) as appropriate for the median 
cluster age characterizing the low mass population. First, we converted spectral types 
derived for the NGC 1333 objects to temperatures using empirical relations derived 
from \citet{da02} and \citet{le96}. The typical uncertainties in temperatures
due to uncertainties in spectral types were $\approx$ 200K. Luminosities were 
derived from the de-reddened J-band magnitudes. To determine M$_J$ we first
calculated visual extinctions using the standard relation 
A$_v$=9.09x[(J-H)$_{obs}$ - (J-H)$_o$] \citep{co81} where (J-H)$_o$
is the intrinsic color for a given spectral type. The values
of A$_v$ derived for NGC 1333 sources lie between 
0 and 14 mag, with 6 sources having an extinction higher than 5 mag. Values for 
log(L$_{bol}$/L$_\odot$) were calculated from the de-reddened
J-band magnitudes, after bolometric correction, as 
log(L$_{bol}$/L$_\odot$) = 1.89 - 0.4 M$_{bol}$. We adopt intrinsic colors and 
bolometric corrections following \citet{wil03}. Typical errors in 
log(L$_{bol}$/L$_\odot$) are $\pm$ 0.2 dex 
and are derived based on errors in the J-band magnitude, the distance modulus
and the bolometric correction. The H-R Diagram is presented in Figure 5.
The luminosities and temperatures for the NGC 1333 objects with spectral types 
are listed in Table 4. 
The positions of the brown dwarfs are compared to the theoretical tracks 
and isochrones from the
models of DM98. In addition to our sample, objects in NGC 1333 for which unique spectral 
types were derived in \citet{wil03} are plotted on the H-R diagram. 
From these objects we estimated the cluster to be younger than 1 Myr and 
chose the DM98 0.3 Myr tracks as appropriate for the median cluster age. It is noteworthy that 
these ages are relative ages as absolute ages are model dependent. The objects in 
the HR Diagram show a scatter around the median age with an overall trend to younger ages for 
lower mass objects. The scatter is due in part to errors in the luminosities of the objects 
introduced by errors in the photometry and dereddening. 
The overall trend towards younger ages for lower mass objects is an effect often 
observed in HR Diagrams of young stellar clusters. This is most likely an effect of 
imperfect PMS models and not a real effect of an age-mass relation for the cluster.
The median age of 0.3 Myr for our H-R diagram corresponds to the break point of 
the mass bins for our analysis (0.1 M$_{\odot}$), thus ensuring that we are 
using the correct mass-luminosity relationship at that mass. 

Four objects exhibit unusual positions in the H-R Diagram, with ages much
older than 1 Myr, appearing under-luminous compared to their PMS counterparts. 
Two of these objects have spectral 
types earlier than M0, meaning they show no water band absorption feature 
in their spectra. These two objects could be background stars. 
The other two objects, both of spectral type M4 also appear fainter than 
expected, but could still be cluster members. Notice also 3 objects
from \citet{wil03} which lie below the expected locus for the majority 
of the PMS objects. 
These objects are specifically addressed in \citet{wil03}, but the following
reasoning applies to them as well. 
The sources could have an infrared excess, causing a dilution of 
the water absorption bands, leading us to assign a spectral type that 
is too early. They could also exhibit unresolved scattered
light due to a surrounding disk, which would result in bluer colors and could 
cause us to underestimate
extinction and thus luminosity. There is evidence that disks around young
brown dwarfs are ubiquitous. \citet{ln03} for example found that 77$\%$ of a sample
of young brown dwarfs in IC348 and Taurus showed evidence of a circumstellar 
disk. To establish whether any of these objects show signs of a disk and thus
confirming their youth, we attempt to assess whether they show an infrared excess in the 
next section. This
method is described in detail in \citet{wgm99}. Since we cannot rule out 
either solution at this point it is impossible to 
assign a mass to the two M4 sources or an age if they are not cluster members. 
For these reasons all four sources have been excluded from the mass function analysis. 
It is possible that other objects in our sample also possess circumstellar disks. 
However, for most objects we can expect the effect of the disks on spectral typing to 
be small (\citet{mey97}; \citet{muz03}). 
Even if objects show a strong infrared excess only those with special viewing geometry 
would be dominated by scattered light leading us to underestimate the reddening. 
Further observations will be required to test whether these objects suffer from the 
effects mentioned.

\subsection{K-band excess}

For NGC 1333 objects with 2MASS photometry we have estimated the amount of excess emission 
at K, defined as r$_k$=F$_{Kex}$/F$_K$\footnotemark[1]\footnotetext[1]{r$_k$ = F$_{Kex}$/F$_K$ = [(1+r$_\lambda$)(10$^{[(H-K)-(H-K)_o-0.065A_v]/2.5}$)-1]. We have 
assumed that r$_j$=r$_h$=0.}, where F$_{Kex}$ is the flux contribution at K due 
to circumstellar material and F$_K$ the inherent stellar flux at K 
(\citet{wil03}; \citet{wgm99}). 
Values of r$_k$ are listed in Table 4. These estimates are 
lower limits since we have assumed no excess at J- or H-band. 
Two objects, ASR25 and ASR105, show moderate K-band excesses. ASR25 has a peculiar 
position on the H-R Diagram as mentioned above. The measured K-band excess makes it a likely cluster 
member and it may indicate that we have assigned it a spectral type that is too early, 
which would make it a brown dwarf, and that it may possess a disk. 
Most of the other objects show little or no excess emission. It is possible that the lowest 
mass objects possess disks which are too cool to cause an excess in the K-band as inner disk temperature 
is expected to be correlated with stellar luminosity (\citet{pas03}).

\subsection{Color-Magnitude Diagram and Extinction-Limited Sample}

In Figure 6 we present a CMD for the cluster. Also
shown is the 0.3 Myr isochrone of DM98, which we adopted based on the H-R
diagram discussed above. The isochrone was converted to the CMD by using a 
set of colors and bolometric corrections, with colors
taken from \citet{le92}, while bolometric corrections were adopted from
\citet{le96}, similar to \citet{wgm99}. All but 3 out of 25 objects in the CMD
 appear to lie below the hydrogen burning limit for a cluster
age of 0.3 Myr. This number changes to 5 if we instead assume a cluster
age of 1 Myr. The photometry, isochrone and reddening vector \citep{co81} 
are all presented in the CIT system and the completeness limit discussed 
above is also shown. 
For the age derived above we define an unbiased sample
of low-mass objects in order to constrain the IMF for the cluster. 
We begin by creating an extinction-limited subsample,
to ensure that we are sampling the stellar population uniformly and not 
over-representing more luminous (massive) deeply embedded objects. The extinction 
limits used are A$_v$ $\leq$ 21.1 mag and $\leq$ 18.3 mag for the 0.3 Myr 
and 1 Myr DM98 isochrones respectively,
corresponding to a mass range of 0.02 M$_\odot$ - 0.1 M$_\odot$.
 For both isochrones this led to a total sample of 13 (17-4) objects, 
excluding the 4 objects mentioned in section 3.2. 

We are sensitive to a mass range lower than 20 M$_{Jup}$, albeit for a smaller
range of extinctions. However, the DM98 tracks do not extend below this mass 
limit. Thus no objects lower than 20 M$_{Jup}$ were considered for the cluster
mass function. This means that compared to the survey of \citet{wil03} we probe 
to a higher A$_v$ but not to a lower mass range.
There are six objects lying in the region below 20 M$_{Jup}$ for an assumed age 
of 0.3 Myr, two of them very close to the 90\% completeness limit. 
One of them, ASR 28b, is a suspected companion to a brown dwarf primary 
and will be dealt with in more
detail in Section 4.3. Another object has been previously classified as ASR 16 and 
appears to have extended nebulosity associated with it. 
 There are tracks available that extend below 20 M$_{Jup}$,
 such as \citet{bar98} and \citet{bu97}. However the NGC 1333 objects lie above the youngest 
tracks available for both sets of models (1 Myr) in the H-R Diagram.
 Thus these tracks are most likely not appropriate for NGC 1333, making the DM98 tracks the 
most appropriate choice for NGC 1333, with a median age of 0.3 Myr. Typical errors in 
assigning masses from different tracks are approximately a factor of 2 \citep{wh03} 
although this is only well tested down to $\approx$ 0.3 M$_\odot$. 

It is also worth noting 
that while the rapid drop in the number of objects below 20 M$_{Jup}$ looks striking, a 
fairly large area in the CMD in that region is covered by a small range in mass namely 
between 5 - 20 M$_{Jup}$, with the masses defined according to tracks by \citet{bar98} and \citet{bu97}. 
It is however curious that the objects we do detect seem clustered 
towards the lower end of this mass range, which is close to our 90\% completion limit. We 
detect no objects between 10 - 20 M$_{Jup}$ with low A$_v$. Due to our small sample size and 
survey area this does not allow us to rule out that such objects exist in the cloud. It is 
however suggestive that NGC 1333 may lack objects below 20 M$_{Jup}$. 
   
It is possible that at least some of the objects are background contaminants. We have tried 
to assess the amount of background contamination by adjusting the number counts measured 
by \citet{luc03} in the Hubble Deep Field South (HDF-S) using F160W. Using their estimates 
for number counts one expects 1.7 +0.4/-1.1 objects between 19$^m$ - 21$^m$ in our survey. However, the HDF-S
points out of the Galaxy and thus this number should be a lower limit since NGC 1333 lies closer to 
the Galactic plane. On the other hand our NGC 1333 fields lie in front of a large amount 
of extinction, which will reduce the number of contaminants.    

\section{Discussion}

\subsection{Cluster IMF}

To constrain the slope of the IMF down to the lowest masses in the cluster we 
derive the ratio of very low-mass
stars (0.076 - 0.1 M$_\odot$) to brown dwarfs (0.02 - 0.076 M$_\odot$) in our survey.
 We calculate this ratio to be 
$R = 3/10 = 0.30 \pm 0.20$ assuming a 0.3 Myr isochrone, with the errors computed due to 
Poisson statistics. For a 1 Myr isochrone this ratio increases to R = 5/8 = 0.625 $\pm$ 0.356.
Both cluster ages give ratios consistent with having been drawn from 
a \citet{ch03} system IMF, which gives a most likely ratio (mode) of R = 0.30. This 
corresponds to a slope of dN/dM $\propto$ M$^{-1.1}$ over this range. For the older 
isochrone there are a larger number of low-mass 
stars but still more brown dwarfs due to the hydrogen burning limit shifting to a fainter 
magnitude for a higher cluster age. Both ratios are lower limits, since the low-mass bin has objects for which 
we do not have spectra and thus contamination by field stars may lead us to 
overestimate the number of brown dwarfs relative to low mass stars in the sample. Above 
0.076 M$_\odot$ all objects but one have assigned spectral types which allow us to assess 
their cluster membership. The object without a spectral type is ASR9a which is a 
previously identified cluster member \citep{as94} which we have resolved to be a binary. 
Below 0.076 M$_\odot$ however, apart from ASR 9b, there are four objects without 
spectral types, which could be possible field stars. 
Thus these ratios imply upper limits for the slope of the cluster mass spectrum 
below 0.1 M$_\odot$ of $\alpha$ $\leq$ 1.1 and 
$\alpha$ $\leq$ 0.55 for 0.3 Myr and 1 Myr isochrones respectively, where 
dN/dM $\propto$ M$^{-\alpha}$ and $\alpha$ = 2.35 is the Salpeter slope.
   
Where does our result stand with regard to comparable clusters and studies? 
\citet{wil03} have explored the mass spectrum of NGC 1333 using an 
extinction limited sample down to 0.04 M$_\odot$ 
finding the ratio of sub-stellar (0.04 - 0.1 M$_\odot$) to stellar objects 
(0.1 - 1 M$_\odot$)
to be R$_{SS}$ = 1.11 +0.8/-0.4. They used this to estimate an upper limit
for the slope of the lower end of the mass function of $\alpha$ $\leq$ 1.6.
This compares well with slopes in the solar neighborhood below the
hydrogen-burning limit, which \citet{rei98} found to be 1 $<$ $\alpha$ $<$ 2.
\citet{al05} have further attempted to constrain the shape of the field star IMF 
below 0.1 M$_\odot$ and found -0.6 $<$ $\alpha$ $<$ 0.6 with 60\% confidence and 
a best fit of $\alpha$ = 0.3.

To derive a broader estimate of the cluster mass function we attempted to combine 
this survey with that of \citet{wil03} to make the cluster mass function more easily 
comparable to ratios published for other regions. We combined all objects 
between 0.1 - 1 M$_\odot$ from \citet{wil03} with an extinction limit of A$_v$ $\leq$ 
12.8$^m$ with the HST objects between 0.03 - 0.1 M$_\odot$ with an extinction limit of A$_v$ 
$\leq$ 21.1$^m$, scaling the number of HST objects by the ratios of the survey areas, which is 
26.3, as well as the extinctions, which is 0.61. We chose not to include the objects below 
0.1 M$_\odot$ from \citet{wil03} as that survey was estimated to have a large field 
star contamination below 0.1 M$_\odot$. We compute this ratio as 
R$_1$ = N(0.1 - 1 M$_\odot$)/N(0.03 - 0.1 M$_\odot$) = 0.14 $\pm$ 0.047 with errors again due to Poisson statistics, using the same cluster age 
and models for both surveys. 
This ratio is abnormally low when compared to other published 
results in this mass range even within the errors. We have recomputed this ratio for 5
different young star clusters based on previously published data sets and results range 
between $\approx$ 3 - 6. All surveys were either extinction limited surveys 
or were normalized to make them as close to extinction limited surveys as possible. 
See Table 5 for computed results and errors.
 
We then tried to compensate for possible differences in stellar density between the 
two surveys by introducing a normalization factor 
in an overlapping mass range between 0.07 - 0.1 M$_\odot$. This normalization factor 
was computed to be 3.5 $\pm$ 2.0. This increases the ratio to R$_2$ = 0.64 $\pm$ 0.38, 
which is higher but still significantly lower than the result for any other cluster. 
This survey was centered on known brown dwarf candidates. Could this bias our result 
towards an artificially high number of brown dwarfs? Each of our fields was centered 
on one brown dwarf candidate, with S3A and S3B including the same candidate but offset 
from each other. If we remove these three candidates from our ratio, the normalized ratio 
increases to R$_3$ = 0.83 $\pm$ 0.41, which is consistent with the ratio previously 
determined. Thus the inclusion of these objects does not fundamentally change our results.

Due to the expected field star contamination in \citet{wil03} both 
ratios are lower limits. However, 
if the observed density of brown dwarfs is constant over the molecular core previously 
surveyed at near-infrared wavelengths, this would imply that brown dwarfs outnumber 
low-mass stars in the cloud.  
This may be some indication that brown dwarfs form in a 
clustered environment within the cloud or are spatially segregated from stars due to 
dynamical evolution. 

\subsection{Pre-Main Sequence Activity}
\citet{ge02} have conducted a Chandra X-Ray study of young stellar objects in NGC 1333.
Two objects in our study, ASR8 and ASR24, show signs of x-ray activity. 
Both are spectroscopically confirmed brown dwarfs and have virtually no 
extinction. This means that 2 of our 8 (excluding the 4 sources discussion in Sec. 3.2) 
or 25\% of our spectroscopically confirmed brown dwarfs are detected in the X-Ray. 
However, most of the other spectroscopic brown dwarfs in our survey have higher extinctions, 
decreasing the chance of detecting existing x-ray activity \citep{pr05}. 

One of the objects in our survey, ASR 16, appears to have extended nebulosity 
associated with it, which is detected only in the F160W image. 
We were not able to obtain a spectrum for the object or the associated nebulosity. 
Extended emission in the infra-red can be associated with emission-line Herbig-Haro 
objects or scattered light envelopes. Since the nebulosity is seen only in the H-band 
and not at J, contrary to expectations from scattered light models \citep{ken97} we 
explore the possibility of an Herbig-Haro outflow associated with this brown dwarf 
candidate in NGC 1333. 
Since the nebulosity emission is bright in the 
F160W filter which covers 1.4 - 1.8 $\micron$ it seems that the emission is likely due 
to [Fe II] at 1.644 $\micron$. 
Such outflows are thought to be associated with accretion onto the central object. \citet{muz05} find signs 
of accretion are common in the spectra of brown dwarfs. Outflows have also recently been detected 
around brown dwarfs \citep{whe05}. We tentatively identify this as a candidate outflow around the 
brown dwarf ASR 16.  

\subsection{Binaries}
There are two sources in our fields which have possible close companions. One was
previously identified as ASR 9, which is now resolved into ASR9a and ASR9b, with a separation of 
0.8''. The other
is ASR28a, which we have determined has a close companion ASR 28b.
ASR28a is one of our brown dwarf candidates with a spectral type of M7. 
ASR28a and ASR28b have a separation of 1.2'' ($\approx$ 350 AU at the distance of NGC 1333) 
and ASR28b is 4.2 magnitudes fainter in H. If ASR28b is indeed a cluster member 
its colors and apparent J-band magnitude (assuming the same A$_v$ as ASR 28a) suggest a 
temperature of $\approx$ 1800 K (spectral 
type L4 \citep{da02}) and a corresponding mass of 5 M$_{Jup}$ \citep{bu97} (with an approximate 
mass ratio of 0.15). 
If confirmed this object would be very similar to the reported planetary mass 
companion of 2MASS1207-3923 (hereafter 2MASS1207).  
ASR28b is too faint to extract a spectrum from this survey and J- and H-band photometry are 
not enough to rule out the possibility that it is a background object earlier than M4. 
To ascertain that it is indeed a cluster member we hope to obtain a J- and H-band 
spectrum of ASR28b in the near future. 

\section{Summary and Conclusions}

We present NICMOS photometry and spectroscopy for 6 regions in NGC 1333 centered 
on previously detected brown dwarfs from \citet{wil03}. We also analyze spectra of 
brown dwarf standards from this program and that of \citet{na00}. Our results are as follows:

\begin{itemize}
\item We detect a total of 27 sources at H and 25 sources at J down to M$_J$ $\leq$ 21$^m$ and 
M$_H$ $\leq$ 20.$5^m$ over a region of 4 x 51.2'' x 51.2'' square arcseconds. 
\item Following \citet{wil03} and \citet{wgm99} we develop a reddening-independent 
water band index to estimate spectral types within one subclass from SNR $\geq$ 30 
NICMOS G141 spectra. With this method we derive spectral types for 14 sources in NGC 1333 
from $\leq$M0 - M8. Seven objects have spectral types $\geq$ M6 suggesting they are young brown dwarfs. 
While we were sensitive to lower mass objects than \citet{wil03}, we detected only objects of comparable mass 
but with much higher extinction. 
\item Using the CMD and an assumed age of 0.3 Myr for the cluster we define an extinction 
limited sample of 13 sources between 0.02 - 0.1 M$_\odot$ with A$_v$ $\leq$ 21.1$^m$. 
Although we are sensitive to lower A$_v$ objects between 10 - 20 M$_{Jup}$, none were 
detected. Though we are limited by the small number of objects in our survey, this 
may be indicative of a lack of objects below 20 M$_{Jup}$ in the cloud. 
\item We compute the ratio of low-mass stars to brown dwarfs in the cluster as R = N(0.076 - 0.1
M$_\odot$)/(0.02 - 0.076 M$_\odot$) = 0.30 $\pm$ 0.20, which is consistent with having 
been drawn from a field-star IMF \citep{ch03}. We also compute the ratio of 
stars to low-mass objects in the cluster in conjunction with the survey by \citet{wil03} as 
R= N(0.1 - 1M$_\odot$)/(0.03 - 0.1M$_\odot$) = 0.64 $\pm$ 0.38, which lies below 
published results for other regions. This may indicate that brown dwarfs form in 
a segregated environment compared to stars in this cluster.
\item Our survey includes several unusual objects, including a possible companion to 
a spectroscopically confirmed brown dwarf ASR 28, which if it is a cluster member would 
have a mass of 5 M$_{Jup}$. In addition one of the brown dwarfs (ASR 16) might have an outflow associated with it, 
which could be confirmed through follow-up spectroscopy.
\end{itemize}

This work was supported 
by a Cottrell Scholar's Award to MRM from the Research Corporation and NASA grant 
HST13-9846. We would like to thank Morten Andersen for sharing data in advance of publication as well 
as helpful suggestions. 
We thank Wilson Liu and Wolfram Freudling for assistance with data reduction and 
Angela Benoist for preliminary work on spectral reduction.

\clearpage

\begin{figure}
\includegraphics[angle=0]{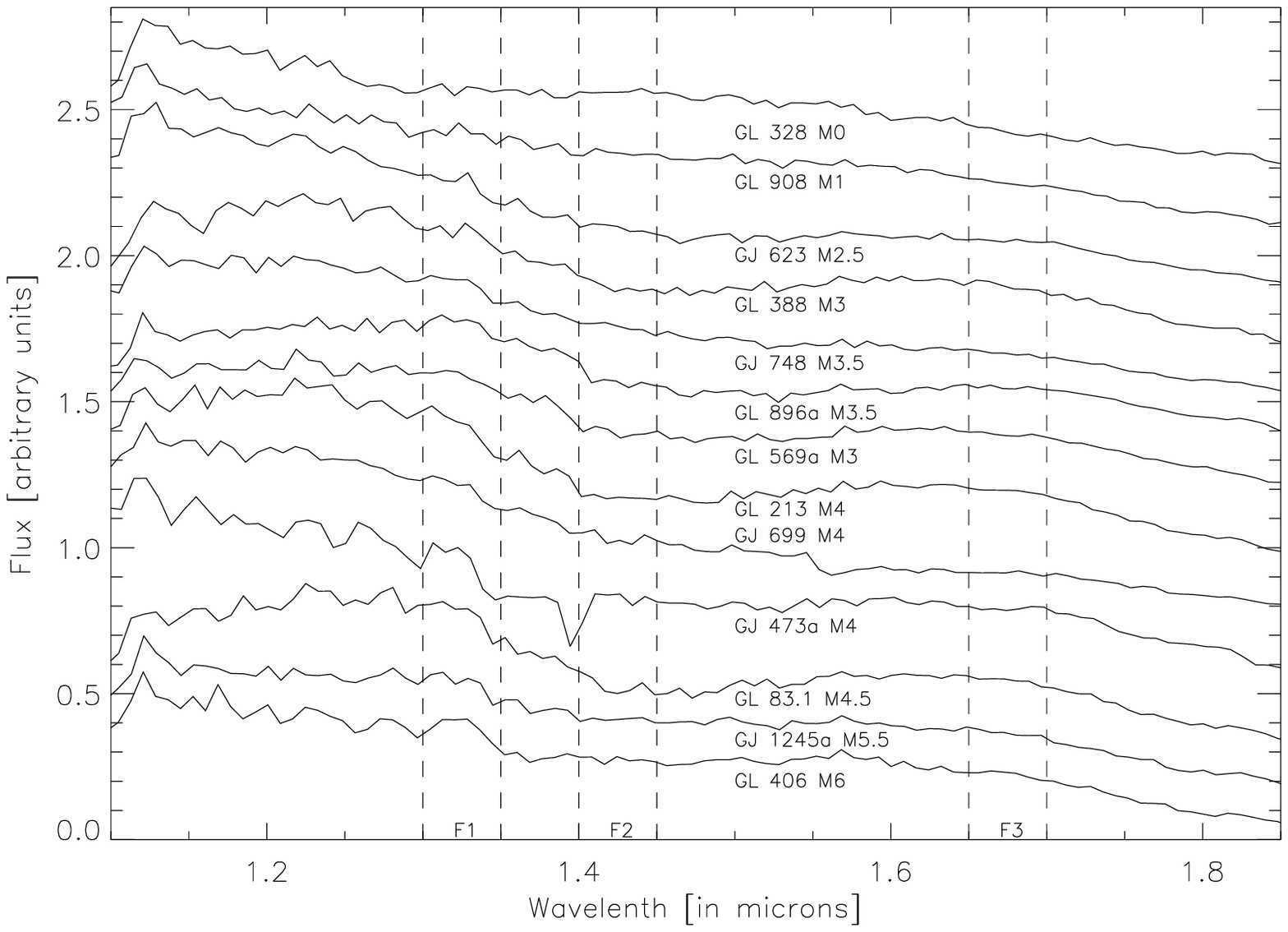}
\caption{NICMOS G141 spectra of field dwarf standards from M0 - M6.}
\end{figure}

\clearpage

\begin{figure}
\includegraphics[angle=0]{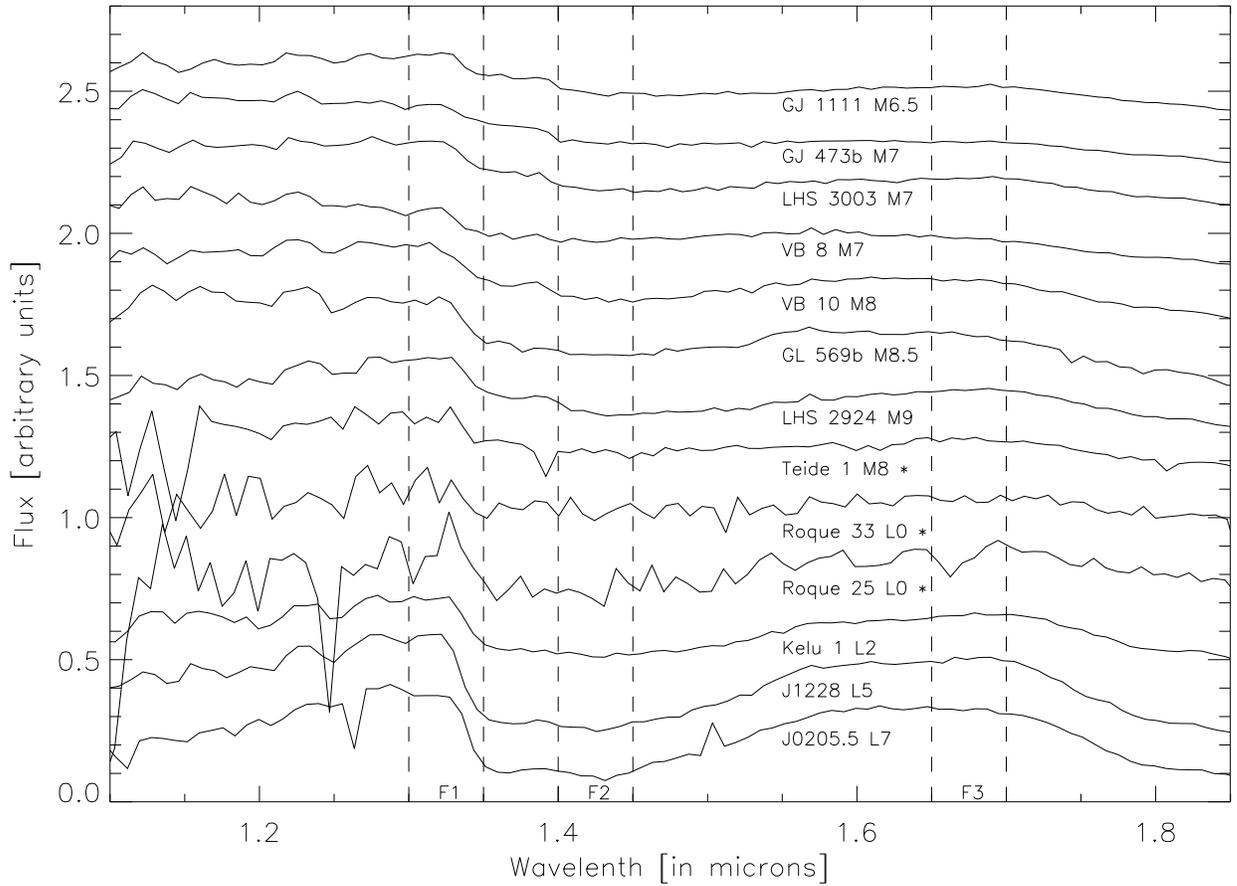}
\caption{NICMOS G141 spectra of field dwarf standards as well as younger brown dwarfs 
confirmed through visible spectroscopy between M6.5 - L7. The younger brown dwarfs are denoted 
by an asterisk.}
\end{figure}

\clearpage

\begin{figure}
\includegraphics[angle=0]{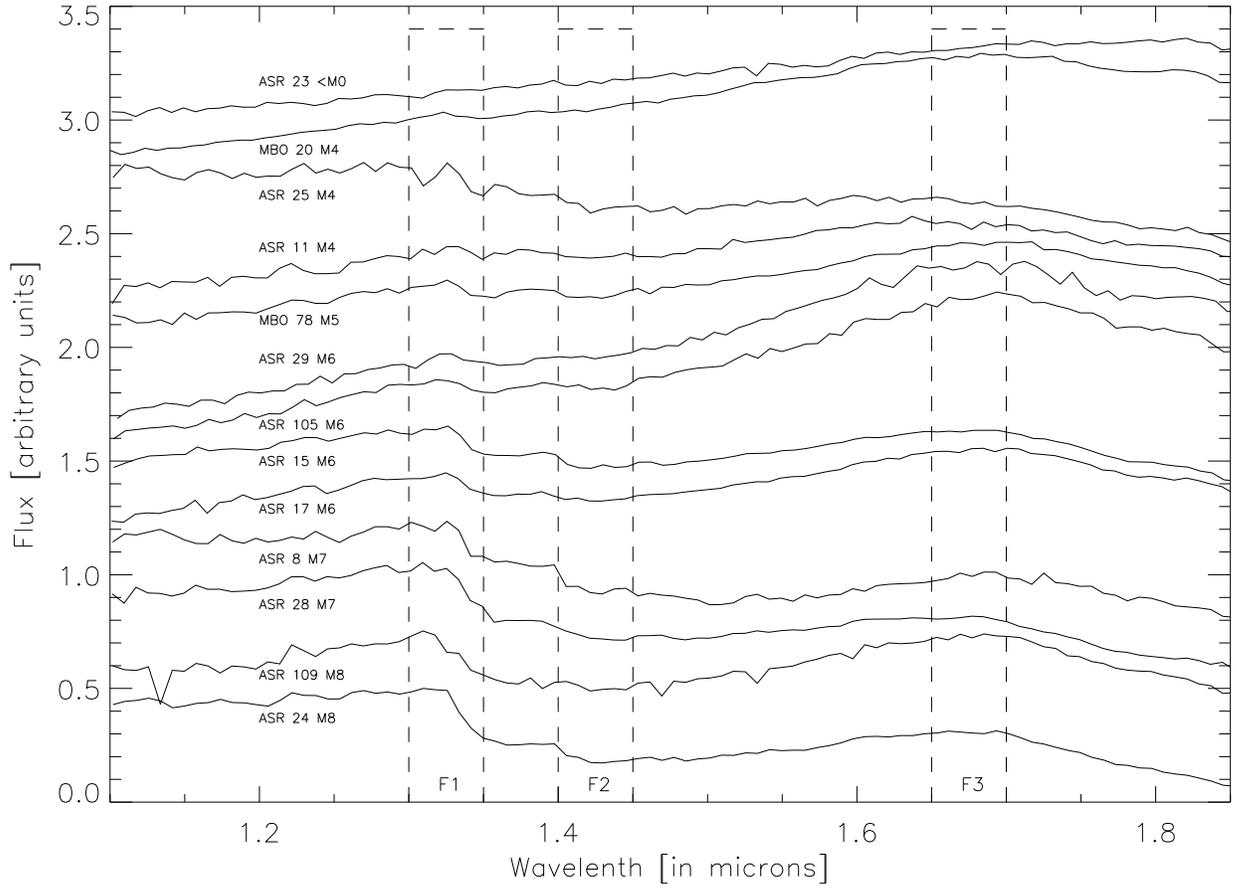}
\caption{NICMOS G141 spectra for candidate brown dwarfs located in NGC 1333, arranged from 
earliest to latest spectral type.}
\end{figure}

\clearpage

\begin{figure}
\plotone{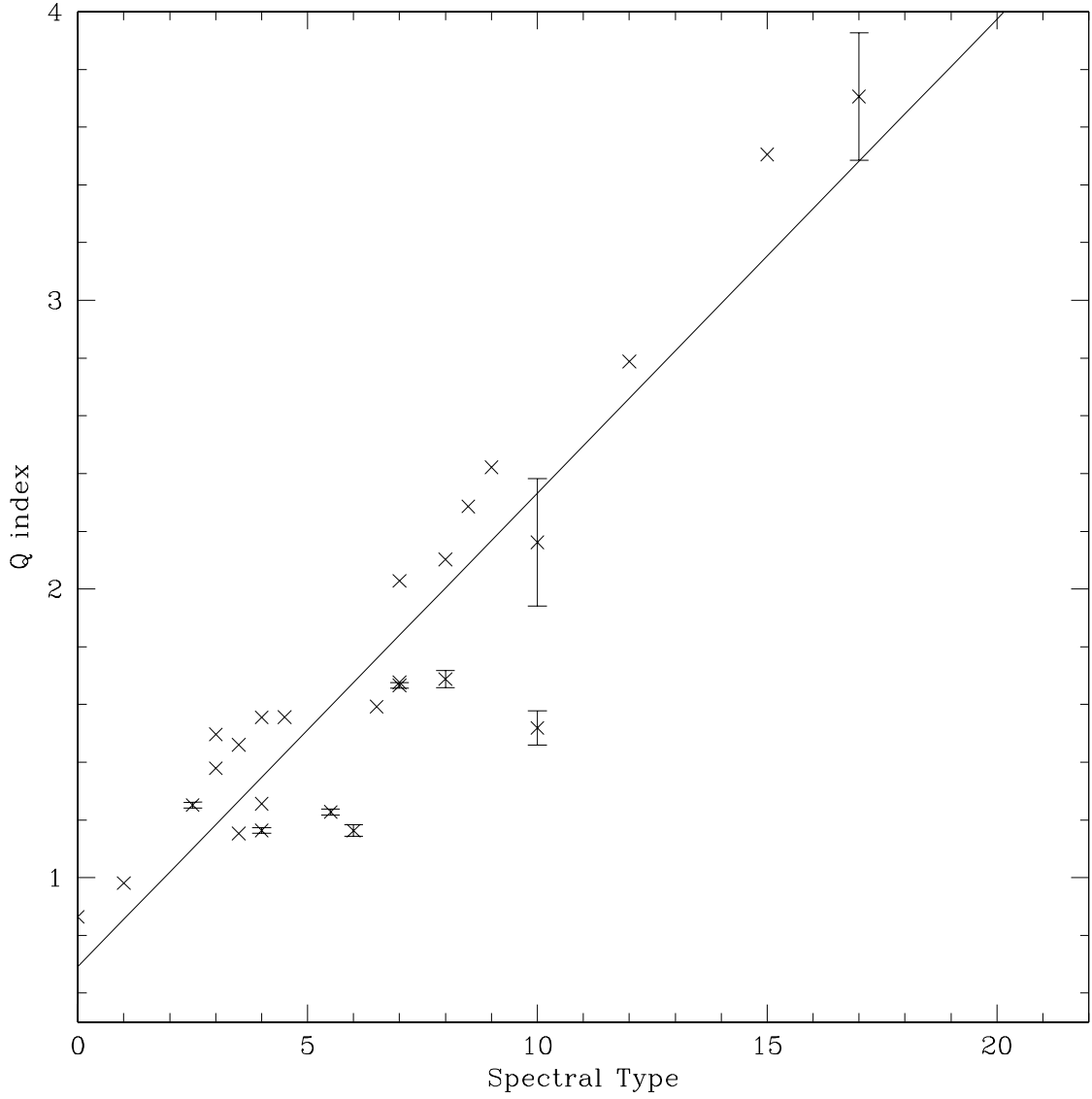}
\caption{Plot of Q vs. spectral type showing the fit derived for the old field dwarfs and young brown dwarfs.}
\end{figure}

\clearpage

\begin{figure}
\plotone{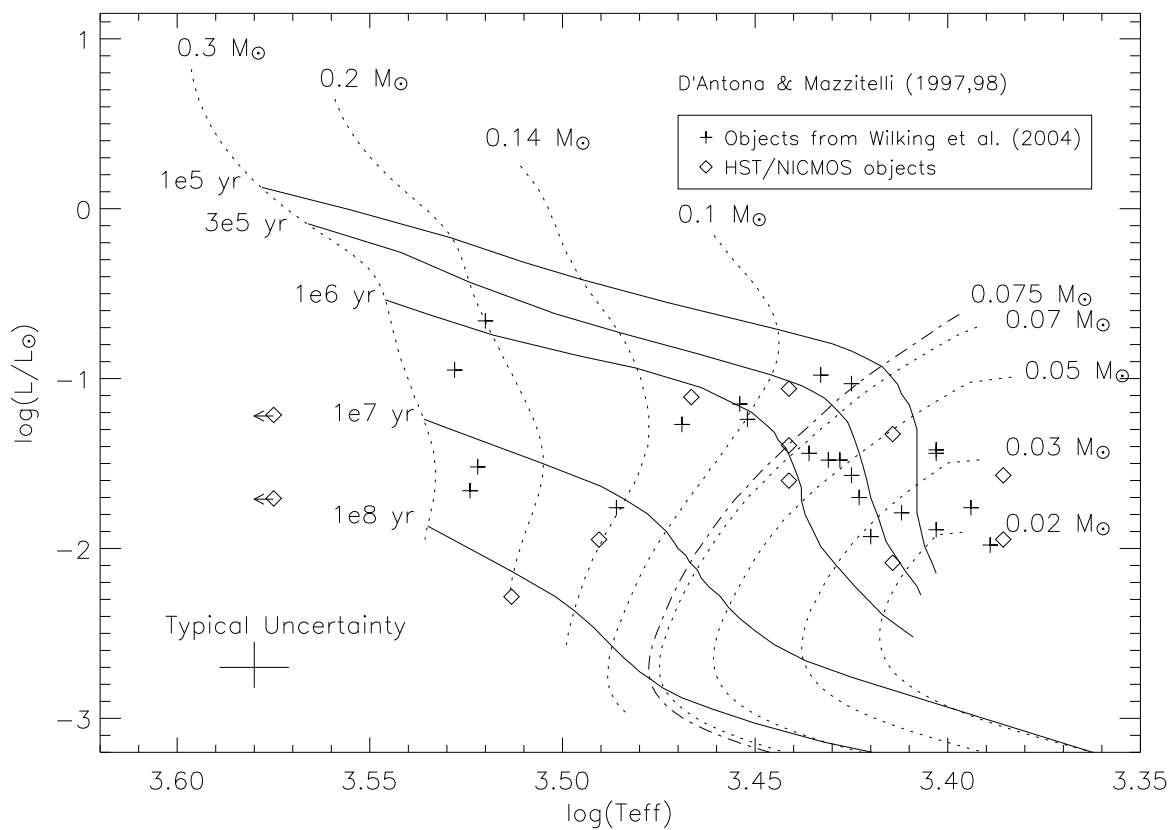}
\caption{H-R Diagram of the NGC 1333 objects from our survey overplotted on
DM98 isochrones and tracks as well as sources from \citet{wil03}. For sources in common 
between the two studies we defer to this study. Four objects show unusual positions in the H-R Diagram. 
They are commented upon in detail in section 3.2.}
\end{figure}

\clearpage

\begin{figure}
\plotone{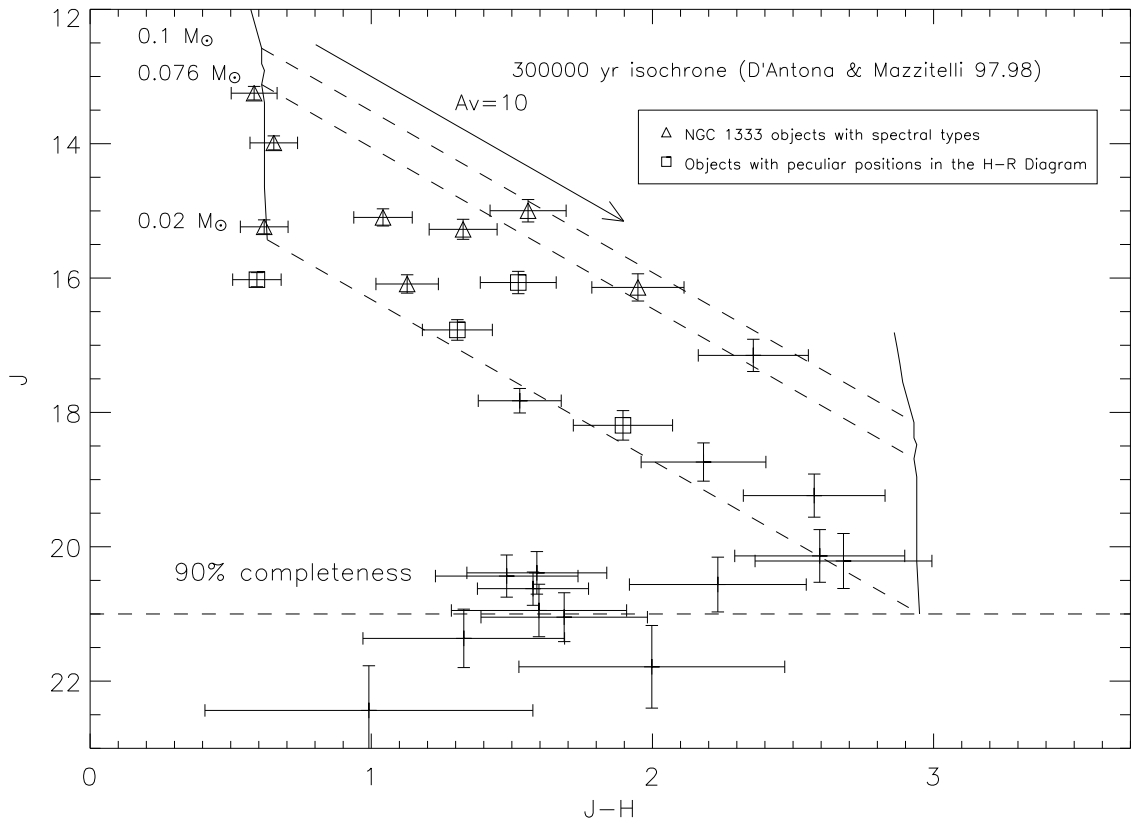}
\caption{NICMOS photometry of sources detected in NGC 1333 transformed into the 
CIT system together with a 0.3 Myr \citep{dm97} isochrone and the reddening 
vector also in the CIT system. The 90\% completeness
limit, derived from artificial star tests, is also shown. The objects with derived spectral types are 
marked.}
\end{figure}

\clearpage

\begin{deluxetable}{ccccc}
\tablecaption{Log of Observations}
\tablehead{
\colhead{Obs. Field} &
\colhead{R.A. (J2000.0)} &
\colhead{Dec. (J2000.0)} &
\colhead{F110 exp. time} &
\colhead{F160 exp. time} 
}
\startdata
S1 & 03:28:56.41 & +31:15:33.1 & 766s & 383s \\ 
S2 & 03:29:03.16 & +31:16:58.7 & 766s & 383s \\
S3A & 03:29:11.97 & +31:16:58.2 & 766s & 383s \\
S3B & 03:29:12.38 & +31:17:25.2 & 766s & 383s \\
N2 & 03:29:11.65 & +31:23:19.5 & 510s & 256s \\
N7 & 03:29:04.1 & +31:25:29.5 & 255s & 128s \\
Roque 25 & 03:48:30.6 & +22:44:50 & 256s & 128s \\
Roque 33 & 03:48:49.0 & +24:20:25 & 256s & 128s \\
Teide 1 & 13:05:40.18 & -25:41:06.0 & 256s & 128s \\
J0205.5-1159 & 02:05:29.40 & -11:59:29.7 & 256s & 128s \\
Sigma Ori 60 & 05:39:37.62 & -02:30:45.64 & 256s & 128s \\
Sigma Ori 47/27 & 05:38:16.00 & -02:40:23.80 & 256s & 128s \\
\enddata
\end{deluxetable}

\clearpage

\begin{deluxetable}{ccccccccccc}
\tablecaption{Photometry of sources detected in J- or H-band}
\tablewidth{0pt}
\rotate
\tablehead{
\colhead{Object ID} &
\colhead{R.A. (J2000.0)} &
\colhead{Dec. (J2000.0)} &
\colhead{J} &
\colhead{Jerr} &
\colhead{H} &
\colhead{Herr} &
\colhead{ASR ID\tablenotemark{a}} &
\colhead{Notes}
}
\startdata
S1-1 & 03:28:54.938 & +31:15:29.65 & 16.09 & 0.14 & 14.96 & 0.05 & 109 \\
S1-2 & 03:28:55.334 & +31:15:53.83 & N/A   & N/A  & 20.00 & 0.23 & N/A \\
S1-3 & 03:28:56.020 & +31:15:49.28 & 20.62 & 0.25 & 19.05 & 0.10 & 16 & Possible Outflow \\
S1-4 & 03:28:56.959 & +31:15:50.93 & 15.10 & 0.13 & 14.06 & 0.04 & 15 \\
S1-5 & 03:28:57.162 & +31:15:35.19 & 15.28 & 0.15 & 13.95 & 0.05 & 17 \\	 
S1-6 & 03:28:57.388 & +31:15:36.72 & 17.83 & 0.18 & 16.30 & 0.07 & N/A \\
S1-7 & 03:28:57.990 & +31:15:50.36 & N/A   & N/A  & 21.03 & 0.39 & N/A \\
S1-8 & 03:28:58.058 & +31:15:48.65 & 19.24 & 0.32 & 16.67 & 0.10 & N/A \\
S2-1 & 03:29:01.673 & +31:17:12.25 & 20.44 & 0.31 & 18.96 & 0.16 & N/A \\
S2-2 & 03:29:01.876 & +31:16:43.98 & 21.79 & 0.62 & 19.79 & 0.24 & N/A \\
S2-3 & 03:29:01.929 & +31:16:53.22 & 16.77 & 0.15 & 15.47 & 0.06 & 11 \\
S2-4 & 03:29:02.654 & +31:17:14.84 & 22.44 & 0.66 & 21.45 & 0.48 & N/A \\
S2-5 & 03:29:04.077 & +31:16:59.71 & 20.95 & 0.39 & 19.36 & 0.19 & N/A \\
S2-6 & 03:29:04.114 & +31:17:07.46 & 13.25 & 0.10 & 12.67 & 0.03 & 8 \\
S2-7 & 03:29:04.179 & +31:16:50.93 & 18.74 & 0.28 & 16.56 & 0.09 & 9b \\
S2-8 & 03:29:04.229 & +31:16:50.60 & 17.15 & 0.24 & 14.80 & 0.09 & 9a \\
S2-9 & 03:29:04.713 & +31:16:59.03 & 15.00 & 0.17 & 13.44 & 0.06 & 105 \\
S2-10 & 03:29:05.413 & +31:17:02.57 & 20.56 & 0.41 & 18.33 & 0.14 & N/A \\
S3B-6 & 03:29:10.716 & +31:17:21.34 & 21.36 & 0.44 & 20.04 & 0.25 & N/A \\
S3A-1 & 03:29:10.787 & +31:16:43.99 & 16.07 & 0.17 & 14.55 & 0.06 & 23 \\
S3A-2 & 03:29:11.272 & +31:17:18.85 & 13.95 & 0.10 & 13.30 & 0.04 & 24 & Both in frame S3A \& S3B \\
S3B-3 & 03:29:11.353 & +31:17:17.52 & 13.99 & 0.09 & 13.31 & 0.03 & 24 & Both in frame S3A \& S3B \\ 
S3A-3 & 03:29:12.923 & +31:17:08.34 & 15.94 & 0.11 & 15.38 & 0.04 & 25 & Both in frame S3A \& S3B \\
S3B-4 & 03:29:13.008 & +31:17:07.01 & 16.02 & 0.15 & 15.40 & 0.04 & 25 & Both in frame S3A \& S3B \\ 
S3A-4 & 03:29:13.014 & +31:16:41.12 & 18.19 & 0.22 & 16.30 & 0.08 & 26 \\
S3B-7 & 03:29:13.034 & +31:17:37.34 & 20.39 & 0.32 & 18.80 & 0.13 & 28b \\
S3B-1 & 03:29:13.078 & +31:17:38.24 & 15.24 & 0.10 & 14.62 & 0.04 & 28a \\
S3A-5 & 03:29:13.269 & +31:17:17.35 & 20.21 & 0.41 & 17.54 & 0.12 & 27 & Both in frame S3A \& S3B \\
S3A-6 & 03:29:13.283 & +31:16:45.91 & 23.37 & 2.38 & 19.47 & 0.37 & N/A \\
S3B-5 & 03:29:13.351 & +31:17:16.01 & 20.14 & 0.39 & 17.55 & 0.12 & 27 & Both in frame S3A \& S3B \\
S3B-2 & 03:29:13.655 & +31:17:43.48 & 16.14 & 0.20 & 14.19 & 0.07 & 29 \\
S3A-7 & 03:29:14.285 & +31:17:00.58 & 21.05 & 0.36 & 19.37 & 0.19 & N/A \\

\enddata
\tablenotetext{a}{ASR refers to \citet{as94}}
\end{deluxetable}

\clearpage

\begin{deluxetable}{ccccccc}
\tablecaption{Spectral Standards}
\tablehead{
\colhead{ID} &
\colhead{R.A. (J2000.0)} &
\colhead{Dec. (J2000.0)} &
\colhead{Spectral Type\tablenotemark{a}} &
\colhead{Q\tablenotemark{b}} &
\colhead{Prop ID}
}
\startdata
GL 328 & 08:55:07.62 & +01:32:47.4 & M0 & 0.87 & 7322 \\   	
GL 908 & 23:49:12.53 & +02:24:04.4 & M1 & 0.98 & 7830 \\
GJ 623 & 16:24:09.32 & +48:21:10.5 & M2.5 & 1.25 $\pm$ 0.01 & 7322 \\
GL 388 & 10:19:36.50 & +19:52:10.6 & M3 & 1.38 & 7322 \\
GL 569a & 15:43:02.12 & +26:16:36.0 & M3 & 1.50 & 7322 \\
GJ 748 & 16:24:09.32 & +48:21:10.5 & M3.5 & 1.15 & 7830 \\
GL 896a & 23:31:52.18 & +19:56:14.1 & M3.5 & 1.46 & 7830 \\
GL 213 & 05:42:09.27 & +12:29:21.6 & M4 & 1.56 & 7322 \\
GJ 699 & 17:57:48.50 & +04:41:36.2 & M4 & 1.16 $\pm$ 0.01 & 7322 \\
GJ 473a & 12:33:16.3 & +09:01:26 & M4 & 1.26 & 7830 \\
GL 83.1 & 16:36:21.45 & -02:19:28.5 & M4.5 & 1.56 & 7322 \\ 
GJ 1245a & 19:53:54.48 & +44:24:53.3 & M5.5 & 1.23 $\pm$ 0.01 & 7830 \\
GL 406 & 10:56:28.99 & +07:00:52.0 & M6 & 1.16 $\pm$ 0.02 & 7322 \\
GJ 1111 & 08:29:49.35 & +26:46:33.7 & M6.5 & 1.59 & 7322 \\
GJ 473b & 12:33:19.1 & +09:01:10 & M7 & 1.67 $\pm$ 0.01 & 7830 \\
LHS 3003 & 14:56:38.31 & -28:09:47.4 & M7 & 2.03 & 7322 \\
VB 8 & 16:55:35.29 & -08:23:40.1 & M7 & 1.68 & 7322  \\                                  
VB 10 & 19:16:57.66 & +05:09:00.4 & M8 & 2.10 & 7322 \\
Teide 1 & 13:05:40.18 & -25:41:06.0 & M8 & 1.69 $\pm$ 0.03 & 9846 \\
GL 569b & 15:43:03.8 & +26:15:59 & M8.5 & 2.29 & 7322 \\
LHS 2924 & 14:28:43.33 & +33:10:37.9 & M9 & 2.42 & 7322 \\
Roque 33 & 03:48:49.0 & +24:20:25 & L0 & 1.52 $\pm$ 0.06 & 9846 \\ 
Roque 25 & 03:48:30.6 & +22:44:50 & L0 & 2.16 $\pm$ 0.22 & 9846 \\
Kelu 1 & 13:05:40.18 & -25:41:06.0 & L2 & 2.79 & 7830 \\ 
J1228-1547 & 12:28:15.23 & -15:47:34.2 & L5 & 3.51 & 7830 \\
J0205.5-1159 & 02:05:29.40 & -11:59:29.7 & L7 & 3.71 $\pm$ 0.22 & 9846 \\
\enddata
\tablenotetext{a}{Spectral Types are from \citet{Khm91}, \citet{hks94}, \citet{ma00} and \citet{k99}}
\tablenotetext{b}{Errors for Q are only given for objects with 4 or more spectra}
\end{deluxetable}

\clearpage

\begin{deluxetable}{ccccccccccc}
\tablecaption{Properties of NGC 1333 objects with spectral types}
\tablewidth{0pt}
\rotate
\tablehead{
\colhead{Object Name\tablenotemark{a}} &
\colhead{Spectral Type} &
\colhead{Log(Teff)\tablenotemark{b}} &
\colhead{J-H\tablenotemark{c}} &
\colhead{H-K} &
\colhead{K} &
\colhead{A$_v$\tablenotemark{d}} &
\colhead{Log(L/L$_\odot$)\tablenotemark{e}} &
\colhead{r$_k$\tablenotemark{f}} &
\colhead{Q} &
\colhead{Notes\tablenotemark{g}}
}
\startdata
ASR24 & M8 $\pm$ 0.8 & 3.38 & 0.60 & 0.42 & 12.94 & 0.0 & -1.52 & 0.02 & 2.02 $\pm$ 0.05 & M8.2 \\
ASR25 & M4 $\pm$ 0.6 & 3.51 & 0.70 & 0.70 & 15.33 & 0.7 & -2.31 & 0.42 & 1.36 $\pm$ 0.09 \\
ASR29 & M5 $\pm$ 0.8 & 3.46 & 2.08 & 1.16 & 13.05 & 13.7 & -1.02 & -0.04 & 1.57 $\pm$ 0.07 \\
ASR109 & M8 $\pm$ 0.8 & 3.40 & 1.01 & 0.67 & 13.24 & 3.6 & -1.92 & -0.01 & 1.93 $\pm$ 0.05 \\
ASR11 & M4 $\pm$ 0.6 & 3.50 & 1.15 & 0.58 & 14.89 & 4.8 & -2.05 & 0.00 & 1.27 $\pm$ 0.02 \\
ASR105 & M6 $\pm$ 0.8 & 3.45 & 1.56 & 1.19 & 12.69 & 8.5 & -1.21 & 0.29 & 1.62 $\pm$ 0.07 \\
ASR17 & M6 $\pm$ 0.7 & 3.44 & 1.36 & 0.75 & 13.21 & 4.8 & -1.37 & 0.08 & 1.67 $\pm$ 0.02 & M7.4 \\
ASR26 & $<$M0 $\pm$ 0.7 & 3.58 & 1.59 & 0.85 & 15.54 & 8.8 & -1.83 & 0.18 & 0.90 $\pm$ 0.01 \\
ASR8 & M7 $\pm$ 1.1 & 3.43 & 0.63 & 0.33 & 12.34 & 0.1 & -1.26 & -0.07 & 1.76 $\pm$ 0.17 \\
ASR28 & M7 $\pm$ 0.8 & 3.41 & 0.64 & 0.39 & 14.18 & 0.2 & -2.00 & -0.02 & 1.89 $\pm$ 0.05 & Binary \\
ASR15 & M6 $\pm$ 0.7 & 3.44 & 0.99 & 0.53 & 13.49 & 3.3 & -1.60 & -0.04 & 1.66 $\pm$ 0.03 & M7.4 \\
ASR23 & $<$M0 $\pm$ 0.6 & 3.58 & 1.48 & 0.76 & 13.33 & 7.8 & -1.04 & 0.16 & 0.89 $\pm$ 0.07 \\
MBO20 & M3 $\pm$ 0.6 & 3.51 & 1.75 & 0.88 & 10.91 & 10.9 & -0.17 & -0.05 & 1.19 $\pm$ 0.02 \\
MBO78 & M5 $\pm$ 0.7 & 3.47 & 1.78 & 1.18 & 13.37 & 10.8 & -1.35 & 0.16 & 1.45 $\pm$ 0.05 \\
\enddata
\tablenotetext{a}{ASR designations are from \citet{as94}; MBO designations are from \citet{wil03}}
\tablenotetext{b}{Typical errors are $\approx$ 0.1}
\tablenotetext{c}{2MASS photometry for the objects listed. This was used to derive r$_k$.}
\tablenotetext{d}{Typical errors are $\approx$ 1}
\tablenotetext{e}{Typical errors are $\approx$ 0.15}
\tablenotetext{f}{Typical errors are $\approx$ 0.1}
\tablenotetext{g}{Spectral types derived by \citet{wil03}}
\end{deluxetable}

\clearpage

\begin{deluxetable}{cc}
\tablecaption{Ratios of intermediate-mass stars to low-mass objects}
\tablehead{
\colhead{Region} &
\colhead{N(0.1-1 M$_\odot$)/N(0.03 - 0.1 M$_\odot$)} 
}
\startdata
Pleiades\tablenotemark{a} & 3.3 +2.2/-1.0\\
Chamaeleon\tablenotemark{b} & 5.6 +2.4/-4.5 \\
Mon R2\tablenotemark{c} & 5.3 $\pm$ 3.35 \\
Taurus\tablenotemark{d} & 4.0 $\pm$ 1.2 \\
IC 348\tablenotemark{e} & 3.0 $\pm$ 0.5 \\
Orion\tablenotemark{f} & 3.3 $\pm$ 0.6 \\
NGC 1333 & R$_1$ = 0.14 $\pm$ 0.047 \\
NGC 1333 & R$_2$ = 0.64 $\pm$ 0.38 \\
\enddata
\tablenotetext{a}{Data from \citet{mo03}}
\tablenotetext{b}{Data from \citet{luh04b}}
\tablenotetext{c}{Data from \citet{an06}}
\tablenotetext{d}{Data from \citet{luh04}}
\tablenotetext{e}{Data from \citet{luh03a}}
\tablenotetext{f}{Data from \citet{hi00}}
\end{deluxetable}

\end{document}